\documentclass[12pt]{article}

\usepackage{xcolor}
\usepackage{subfig}
\usepackage{graphicx}
\usepackage{hyperref}

\newtheorem{remark}{Remark}

\newcommand{\rmd}{{\rm d}}
\newcommand{\rme}{{\rm e}}
\newcommand{\rmi}{{\rm i}}

\begin{document}

\title{Amplification of quantum transfer and quantum ratchet}
\author{Sergei V. Kozyrev$^1$\footnote{E-mail: {kozyrev@mi-ras.ru}; ORCID \href{https://orcid.org/0000-0002-5597-5556}{0000-0002-5597-5556};  \href{https://www.mathnet.ru/eng/person11785}{mathnet.ru/eng/person11785}} and Alexander N. Pechen$^{1,2}$\footnote{E-mail: {apechen@gmail.com}; ORCID \href{https://orcid.org/0000-0001-8290-8300}{0000-0001-8290-8300};  \href{http://www.mathnet.ru/eng/person17991}{mathnet.ru/eng/person17991}}}

\date{}

\maketitle

\centerline{$^1$ Steklov Mathematical Institute of Russian Academy of Sciences, }
\centerline{Gubkina str. 8, Moscow 119991, Russian Federation}
\centerline{$^2$ Quantum Engineering Research and Education Center, University of Science and Technology MISIS, }
\centerline{Moscow, 119991, Russian Federation}
	
\begin{abstract}
Amplification of quantum transfer and ratchet--type processes are important for quantum technologies. We also expect that quantum ratchet works in quantum photosynthesis, where possible role of quantum effects is now widely discussed but the underlying dynamical processes are still not clearly known. In this work, we study a model of amplification of quantum transfer and making it directed which we call the quantum ratchet model. The model is based on a special quantum control master equation with dynamics induced by a feedback-type process. The ratchet effect is achieved in the quantum control model with dissipation and sink, where the Hamiltonian depends on vibrations in the energy difference synchronized with transitions between energy levels. A similarity between this model and the model of coherent transport in quantum photosynthesis, where the time dependence of the Hamiltonian arises due to vibrons, is studied. Amplitude and frequency of the oscillating vibron together with the dephasing rate are the parameters of the quantum ratchet which determine its efficiency. We study with which parameters the quantum ratchet minimizes the exction recombination time and show that the experimentally known values of the parameters of the photosynthetic reaction center correspond to values of the parameters of the quantum ratchet which realize a local minimum of the exciton recombination time. We also find another values of the parameters of the quantum ratchet minimizing the exciton recombination time, which corresponds to a twice smaller frequency of the vibron compared to that observed in experiments.
\end{abstract}

\section{Introduction}
	
In this work, we discuss the following question: How to amplify the quantum transfer and make it directed? There are different regimes of the dynamics of a quantum system, distinguished by the value of the coupling constant $\lambda$ in the expansion of the Hamiltonian into the free $H_0$ and interacting $H_I$ parts
$$
H=H_0+\lambda H_I.
$$
	
In the weak coupling regime (i.e., for small $\lambda$), the dynamics can be considered as a perturbation of the free dynamics by the weak interaction. In this regime, upon transition to a new slow time scale effects of dissipation and decoherence do appear and the dynamics can be represented as generated by a dissipative generator in the Gorini--Kossakowski--Lindblad--Sudarshan (GKSL) form~\cite{Davies1976,SpohnACP1978,Stochastic}. In the strong coupling regime, a characteristic effect is the appearance of fast transitions between levels according to the Landau--Zener mechanism~\cite{Landau,Zener}. The most complicated is the regime of intermediate coupling, when the contributions from the free $H_0$ and interacting $H_I$ parts of the Hamiltonian are comparable. In the intermediate coupling regime, analytical estimates of the dynamics are no longer possible; the dynamics can only be studied numerically. In this regime new effects may arise that may be of interest for quantum technologies~\cite{KochEPJ2022}.
	
An important task in quantum control for applications to quantum technologies is to amplify quantum transfer and make it directed, e.g. to achieve a situation when the rate of forward transfer exceeds the rate of reverse transfer. In principle, dissipative transitions have this property, since transitions with a decrease in energy have a higher probability than transitions with an increase in energy (in the classical statistical mechanics, such behavior is described by the Arrhenius formula, and for quantum dissipative generators in the GKSL form a direct generalization takes place). In opposite, tunnel transitions have the same forward and reverse transfer rates. A specially interesting regime is with an intermediate coupling, where the quantum transport contains both dissipative and tunneling contributions, while the energy difference is small and the level of directivity of the dissipative transition is low. In this case, how can we make the transition directional using its tunnel part?
	
This kind of effect can be achieved exploiting the dependence of the Hamiltonian on time. The tunnel transition between some two states $|1\rangle$ and $|2\rangle$ is accompanied by quantum beats --- oscillations of level populations with a period equal to twice the tunnel transition time. In this case, the amplitude and period of the transition depend on the difference in the energies of the transition, the higher the difference in energies, the lower the amplitude of the transition. If the level energy difference is forced to oscillate with the beat frequency for the tunnel transition, then by adjusting the phase of such oscillations one can achieve that for the forward $|1\rangle\rightarrow|2\rangle$ and the reverse $|2\rangle\rightarrow|1\rangle$ transitions the energy differences are different on average, and that makes the transition directed. For example, the forward transition can be amplified and the reverse transition suppressed.
	
In this work we propose to consider such a behavior for the charge separation transition in quantum photosynthesis based on the model previously developed for explaining charge separation process in photosynthesis~\cite{PhysRevA}. Quantum photosynthesis has been studied in many works, e.g.~\cite{Engel,LeeFleming2007,Fleming2009,Scholes2010,FlemingScience2011,FlemingPlenio2011,Cao2012,Jonas2012,TorresJPCB2014,Miller2014,ChenuScholes2015,Brumer2017,Brumer2018,YangCao2020,Novoderezhkin2015,Novoderezhkin2016,Novoderezhkin2017,Novoderezhkin2017_1,Novoderezhkin2019,Novoderezhkin2021}.
In particular, constructive role of incoherent light and environment in light-harvesting energy transfer has been studied, e.g.~\cite{Brumer2017,Brumer2018,YangCao2020}. In this regard,  a constructive significant role of incoherent light (called as {\it incoherent control}) for quantum control and quantum transport in general context based on the scheme of environment-assisted quantum control was shown in ~\cite{PechenRabitz2006,PechenPRA2011} and importance of incoherent light for quantum transport is actively studied now for various quantum systems~\cite{PetruhanovPhotonics2023,PetruhanovJPA2023,MorzhinQIP2023}. Environment-assisted quantum transport is studied also e.g., in~\cite{Rebentrost2009}, etc. Advanced experimental works have been performed including on coherent feedback control of two-dimensional excitons by engineering of the photonic environment~\cite{Rogers2020}. Based on the general properties of open quantum system dynamics for in biological and chemical systems global optimality of fitness landscape was described for chemical~\cite{MooreCS2011,MoorePCCP2011} and biological~\cite{FengCS2012} processes.
	
Coherent quantum control in photosynthetic systems was studied in variety of works including~\cite{VoroninJPCA2011,Caruso2012,HoyerNJP2014,MukamelPRA2012}, where the control was carried out by a laser. In this paper, we consider vibrons as a coherent control acting on the system via measurement-like feedback mechanism. Various aspects of measurement-based and feedback quantum control were studied for example in~\cite{Belavkin1983,Doherty2000,Handel2005,WisemanMilburn2009,Pechen2006,Shuang2008,Barchielli,JohnGough,Pechen2015,Jacobs2007,Rogers2020} and in many more other works. A detailed discussion of directions in quantum control is provided in the review~\cite{KochEPJ2022}. Our work is based on the quantum feedback control by vibrons in quantum photosynthesis as was proposed in~\cite{PhysRevA}. In quantum description of photosynthesis, the two states $|1\rangle$ and $|2\rangle$ correspond to electronic (exciton) states on a special pair of chlorophylls of the photosynthetic reaction center; the $|1\rangle\rightarrow|2\rangle$ transition is the charge separation transition; the dependence of the Hamiltonian on time models the presence in the system of vibrons --- special vibrations of chromophore nuclei. We model vibrons classically via special time dependence of the parameters of the electronic Hamiltonian. Vibrons in this model arise by the semiclassical Franck--Condon mechanism --- when the electronic degrees of freedom of the chromophores are excited, the nuclei of atoms in the chromophores find themselves in a nonequilibrium state and begin to oscillate. With this mechanism of excitation of vibrons, their phase is adjusted to the electronic states which is important for the mechanism described above.
	
Thus, the creation of excitons generates vibrons, which in turn control the dynamics of excitons. Such behavior is described by the model of quantum feedback control, which is expressed by a nonlinear master equations which determines the dynamics of the system --- the expression for the vibron in the master equation depends on the density matrix for the electronic degrees of freedom. Here the quantum feedback control mechanism arises due to Franck--Condon principle in the semiclassical form, and the feedback provides the phase matching necessary for the described effect. This feedback and phase-matching makes the considered mechanism different from simple excitation by lasers or electromagnetic field.
	
We call the described mechanism for the appearance of directionality in quantum transfer due to interaction with vibrons as the {\it quantum ratchet} mechanism since it has a certain similarity with the classical ratchet mechanism as described in section~\ref{Sec:Ratchet}. The low-energy limit of chemical reactions, where both activation and tunneling mechanisms contribute, was considered in~\cite{Goldanskii}. In the present work, a numerical optimization of the parameters of the quantum ratchet (i.e., amplitude and frequency of the oscillating vibron and the dephasing rate) is carried out and it is shown that for the parameters of the quantum ratchet corresponding to the parameters of the photosynthetic reaction center~\cite{Novoderezhkin2016}, a regime of high efficiency and high directionality of quantum transfer is achieved and hence high efficiency of photosynthesis can be explained using the quantum ratchet model described in this paper. In addition to the observed in experiments, we also find another possible optimal values of the parameters of the quantum ratchet which also provide high efficiency and high directionality of the quantum transfer. These optimal values correspond to a twice smaller frequency of the virbon. Establishing the possibility for it to exist or to not exist in experimental situations is an open question. Beyond photosynthesis, it can play the role for optimization of quantum transport.
	
Various models and theoretical mechanisms for quantum ratchets, as well as experimental implementations, have been considered by researchers. Quantum Brownian motion in adiabatically rocked ratchet systems was investigated for a tunneling induced reversal of quantum current~\cite{ReimannPRL1997}. Models of ratchets driven by harmonic and white noise were investigated~\cite{Bartussek1997}. Electron transport in a quantum ratchet based on an asymmetric triangular quantum dot was investigated experimentally and theoretically~\cite{LinkeEPL1998}. Quantum ratchets in application of weak-localization effect in mesoscopic chaotic dots were proposed~\cite{Ishio2001}. Quantum chaotic dissipative ratchets were studied for cold atoms and Bose-Einstein condensates in periodically flashed optical lattices~\cite{CasatiJPhys2007}. Quantum ratchet control based on Landau-Zener transitions between Floquet states with adiabatic and diabatic ramping to control the resulting directed transport was studied~\cite{Morales-MolinaEPL2008}. Quantum dissipative Rashba spin ratchets to generate a finite stationary spin current by applying an unbiased ac driving to a quasi-one-dimensional asymmetric periodic were predicted~\cite{SmirnovPRL2008}. The quantum ratchet effect under the influence of weak dissipation treated within a Floquet-Markov master equation approach was investigated~\cite{DenisovEPL2009}. Tunnel oscillations of a biased double quantum dot were shown to be employed as driving source for a quantum ratchet~\cite{StarkEPL2010}. A feedback-controlled Brownian ratchet operated by a temperature switch was studied~\cite{Zhang2012}. A robust superconducting ratchet device based on topologically frustrated spin ice nanomagnets was designed~\cite{Rollano2019}. A closely related is the model of quantum Maxwell's demon which was described using quantum feedback control~\cite{Jacobs2009} based on the work~\cite{Alicki2004} (a model of classical Maxwell's demon was considered by Smoluchowski~\cite{Smoluchowski} and discussed by Feynman in his lectures~\cite{Feynman}; the demon was based on ratchet which performed separation of molecules).
	
The structure of this work is the following. In section~\ref{Sec:Ratchet}, we provide the proposed model of quantum ratchet which is different from considered before. In section~\ref{Sec:Photo}, we discuss application of this model to quantum photosynthesis. Next section~\ref{Sec:Simulations} contains results of numerical simulations for minimization of excition recombination time in our model. Discussion section~\ref{Sec:Discussion} contains important comments on the model and the numerical results.
	
\section{The model of the quantum ratchet}\label{Sec:Ratchet}
	
The simplest model of quantum transport is described by a two-state quantum system with $2\times 2$ density matrix $\rho(t)$ which satisfies the von Neumann equation with Hamiltonian $H$:
$$
\frac{\rmd}{\rmd t}\rho(t)=-\rmi[H,\rho(t)],\quad H=\left(\begin{array}{cc} E_1 & J \\ J & E_2 \end{array}\right),
$$
where $J$ is (a real) coupling between the states $|1\rangle=\left(\begin{array}{c} 1 \\ 0 \end{array}\right)$ and $|2\rangle=\left(\begin{array}{c} 0 \\ 1 \end{array}\right)$, $E_1$ and $E_2$ are the energies of the states $|1\rangle$ and $|2\rangle$ in the absence of the coupling. Eigenvalues and eigenvectors of the Hamiltonian $H$ are
\begin{eqnarray*}
E_{\pm}&=&E\pm\sqrt{J^2+\Delta^2},\\
|\psi_{\pm}\rangle&=&\frac{1}{\sqrt{2\left(J^2+\Delta^2\pm\Delta\sqrt{J^2+\Delta^2}\right)}}\left(\begin{array}{c} \Delta\pm\sqrt{J^2+\Delta^2} \cr J \end{array}\right).
\end{eqnarray*}
where $E=(E_1+E_{2})/2$ and $\Delta=(E_1-E_{2})/2$.
	
For $E_1\ne E_2$ the evolution does not lead to a complete transition of the wave function between the states $|1\rangle$ and $|2\rangle$. Instead, it leads to quantum beats between these states with the frequency $2\sqrt{J^2+\Delta^2}$, so that the amplitude of the transition behaves as
\begin{equation}\label{Amp}
\langle 2|\rme^{-\rmi tH}|1\rangle=\frac{J}{2\sqrt{J^2+\Delta^2}}e^{-\rmi tE_{+}} \left(1-\rme^{2\rmi t\sqrt{J^2+\Delta^2}}\right).
\end{equation}
Decreasing of $\Delta$ enhances the transfer: transfer is effective when $\Delta<J$, in this case the amplitude of quantum beats can be close to one.
	
Let us discuss in a non-rigorous way how to make the above quantum beats non--symmetric and the corresponding quantum transfer directed. If one would take in~(\ref{Amp}) energy level splitting $\Delta=\Delta(t)$ as depending on $t$, and synchronize this dependence with quantum beats of the transition, then one could make the above amplitude as dependent on the direction of the transition --- for example large for the direct transition $|1\rangle\rightarrow|2\rangle$ and small for the reverse transition $|2\rangle\rightarrow|1\rangle$. In this way quantum control via tuning of time dependence of the Hamiltonian could amplify the state transfer and make it directed. A careful investigation of the non-symmetric tunneling transitions with numerical simulations is performed below. In particular, this mechanism depends crucially of phase matching between quantum beats and vibrons (produced by oscillations of $\Delta$) which are provided by the quantum feedback mechanism. Thus our scheme is different from excitation of quantum states by an external, e.g., electromagnetic, field without induction of a phase-matching between electronic degrees of freedom and external field (in quantum non-linear optics some kind of phase matching could occur by a different way). In our considered below investigation of quantum photosynthesis, the phase matching of the vibrons and quantum beats is provided by the Franck--Condon principle which in semiclassical form correspond to the quantum feedback model. Simple application of an electromagnetic field to a quantum system does not give synchronization with quantum beats and does not provide the ratchet effect.
	
In works on quantum photosynthesis an importance of resonance in energy between the charge separation transition and the vibron was mentioned, e.g.~\cite{Novoderezhkin2015,Novoderezhkin2016,Novoderezhkin2017,Novoderezhkin2017_1}. In our approach, the vibron is given by the formula~(\ref{thevibron}) below and a resonance in frequency between quantum beats and vibrons is important. Resonances in energy and frequency for photosynthetic centers are related due to particular properties of these quantum systems.
	
\begin{figure}[t!]
\center
\includegraphics[width = 0.6\linewidth]{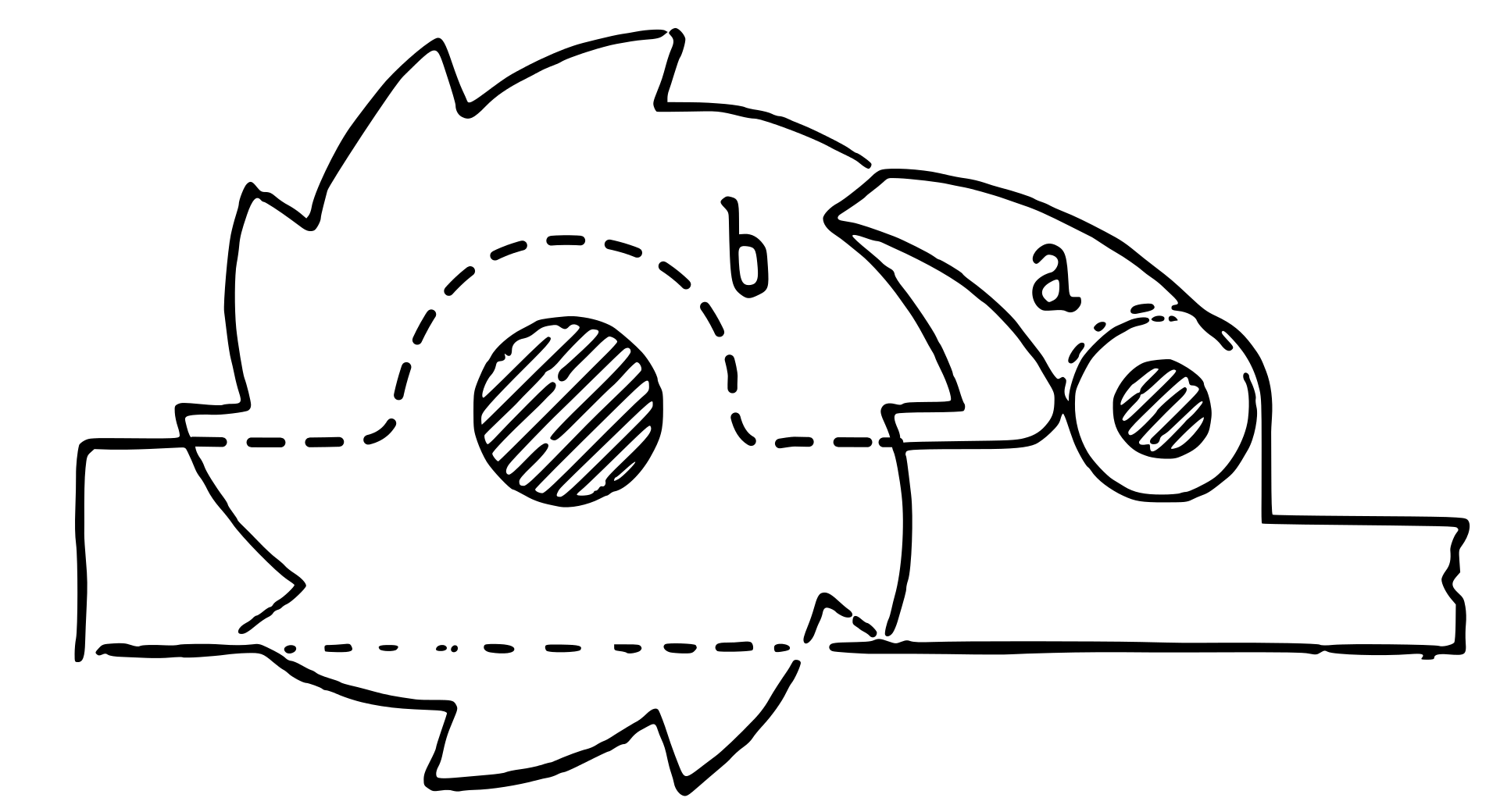}
\caption{Schematic of a ratchet: a) pawl and b) ratchet teeth. This scheme allows for rotation only in the anti-clockwise direction.}
\label{Fig1}
\end{figure}

The above arguments can be compared with the ratchet scheme: ratchet is a round gear with asymmetrical teeth and a pawl which allows for rotations only in one direction as shown on figure~\ref{Fig1} (the picture is taken from~\cite{Klinke}). In quantum ratchet, one can consider directed transitions in a quantum system with state transfer with beats, where oscillations of $\Delta$ in~(\ref{Amp}) serve as an analogue of the asymmetrical teeth in the classical ratchet.

\section{Quantum ratchet in quantum photosynthesis}\label{Sec:Photo}
	
\subsection{Quantum ratchet in an open quantum system}
	
For application of quantum ratchet in quantum photosynthesis, we consider a model of quantum transport in a two-level quantum system with dissipation with master equation
\begin{equation}\label{themaster}
\frac{\rmd}{\rmd t}\rho(t)=-\rmi[H(t),\rho(t)]+{\cal L}(\rho(t))+{\cal S}(\rho(t)),
\end{equation}
Here $H(t)$ is the time dependent Hamiltonian of the two-level system,
\begin{equation}\label{feedback}
H(t)=\left(\begin{array}{cc} E_1+{\bf u}\cdot {\bf q}(t) & J \\ J & E_2+{\bf v}\cdot {\bf q}(t)\end{array}\right),
\end{equation}
where
\begin{equation}\label{thevibron}
{\bf q}(t)={\bf w} \sin (\omega t+\phi).
\end{equation}
is the vibron. This model is invented to describe charge separation in quantum photosynthesis in the presence of vibrons ${\bf q}(t)$ described classically, where ${\bf u}$, ${\bf v}$, ${\bf w}$ are vectors in the space of coordinates of nuclei in chromophores, $\omega$ and $\phi$ are the frequency and the phase of the vibron. The excitation is described by the level $|1\rangle$ of the system. These excitations are created by the absorption of light. The charge separation corresponds to the transition $|1\rangle\rightarrow |2\rangle$.
	
Decoherence and dissipation are described by the GKSL term with vectors $|1\rangle$ and $|2\rangle$,
\begin{equation}\label{theta}
{\cal L}(\rho)=
\gamma^{+}\left(\langle 2|\rho|2\rangle |1\rangle\langle 1|-\frac12
\{\rho,|2\rangle\langle 2|\}\right)+\gamma^{-}\left(\langle 1| \rho |1\rangle |2\rangle\langle 2|-\frac12\{\rho,|1\rangle\langle 1| \}\right).
\end{equation}
Here $\gamma^{+}/\gamma^{-}=\rme^{-\beta(E_1-E_2)}$ (this relation is the quantum analogue of the Arrhenius formula), where $E_1$ and $E_2$ ($E_1>E_2$) are energies of the states $|1\rangle$ and $|2\rangle$, $\beta$ is the inverse temperature, and $\{A,B\}:=AB+BA$ denotes anticommutator.
	
The following term describes sink of excitations:
\begin{equation}\label{sink}
{\cal S}(\rho)=-s_1\langle 1|\rho|1\rangle|1\rangle\langle 1|-s_2\langle 2|\rho|2\rangle|2\rangle\langle 2|,\quad s_1,s_2>0.
\end{equation}
The physical meaning of the first term $-s_1\langle 1 |\rho|1\rangle|1\rangle\langle 1|$ is that it describes the recombination of excitons, while the second term $-s_2\langle 2|\rho|2\rangle|2\rangle\langle 2|$ describes transfer of the electron along the charge transfer chain.
	
To make the process of quantum photosynthesis more efficient, it is necessary to minimize recombination through the first term of the sink operator (\ref{sink}). The exciton recombination is proportional to the effective time the system stays in the state $|1\rangle$, which is
\begin{equation}\label{recombination}
\overline{T}=\int_0^{T_0}\rho_{11}(t)\rmd t,
\end{equation}
where $\rho(t)$ is the solution of Eq.~(\ref{themaster}) with the initial condition $\rho(0)=|1\rangle\langle 1|$ and $T_0$ is some characteristic time period. As discussed in the next section, this time period in Eq.~(\ref{recombination}) should be approximately equal to the period of two quantum beats.
	
Thus we come to the following \emph{problem of quantum control}: minimize the objective~(\ref{recombination}) by tuning parameters of the vibrons ${\bf q}(t)$ in the time dependent Hamiltonian $H(t)$ and the friction parameter $\gamma^{-}$ in the dissipative term (\ref{theta}) to prevent recombination of excitons and make transfer faster and more directed.
	
\begin{remark} The sink term (\ref{sink}) can be considered as a truncation of some GKSL operator in the following sense. Let us apply a projection to a GKSL term which describes transitions between some pair of orthogonal levels  $|A\rangle$ and $|B\rangle$:
$$
|A\rangle\langle A|\left(\langle A|\rho|A\rangle |B\rangle\langle B|-\frac12
\{\rho,|A\rangle\langle A|\}\right)|A\rangle\langle A|=-\langle A|\rho|A\rangle  |A\rangle\langle A|.
$$
This gives exactly a sink term as in (\ref{sink}). Thus the sink term can be considered as a result of a truncation (projection) in a more complex open quantum system where sink is described by dissipative transitions to some ancillary states.
\end{remark}

\subsection{Quantum feedback model}
	
As mentioned above, the vibrons are generated according to the Franck--Condon principle --- when excitons are produced they initiate nuclei vibrations (vibrons) which modulate transitions between the electronic levels. Semiclassical (i.e., when the vibrons are described classically) Franck--Condon principle can be described by a quantum feedback model as a generalization of~(\ref{themaster}), where the Hamiltonian becomes depending on the system density matrix
\begin{equation}\label{themaster1}
\frac{\rmd}{\rmd t}\rho(t)=-\rmi[H(\rho(t),t),\rho(t)]+{\cal L}(\rho(t))+{\cal S}(\rho(t)).
\end{equation}
Here the {\it state-dependent} Hamiltonian
\begin{equation}\label{feedback1}
H(\rho(t),t)=\left(\begin{array}{cc} E_1+{\bf u}\cdot {\bf q}(\rho(t),t) & J \\ J & E_2+{\bf v}\cdot {\bf q}(\rho(t),t)\end{array}\right)
\end{equation}
depends on the system state $\rho(t)$ via the vibrons (i.e. the vibrons are dependent on $\rho(t)$)
\begin{equation}\label{thevibron1}
{\bf q}(\rho(t),t)=\left(\rho_{11}(t)+\rho_{22}(t)\right){\bf w} \sin (\omega t).
\end{equation}
Generators ${\cal L}$ and ${\cal S}$ are as above.
	
Let us note that trace $\rho_{11}(t)+\rho_{22}(t)$ of the density matrix $\rho(t)$ is not conserved by the dynamics because of the presence of the sink term, that is, $\rho(t)$ is only a part of the full density matrix which also contains contributions from the ancillary states.
	
Equations with quadratic nonlinearity and feedback also arise in the semiclassical theory of laser \cite{Haken}, where the laser mode is described classically while the electronic states of atoms are described in a quantum way.

\subsection{Quantum ratchet}
	
The following mechanism is proposed to enhance the quantum transfer in the model (\ref{themaster}) and to make it directional so that speed of the direct transition $|1\rangle\rightarrow|2\rangle$ should exceed speed of the reverse transition $|2\rangle\rightarrow|1\rangle$. Such a mechanism (quantum ratchet) includes two elements:
	
1) Resonance of the transition with the vibron. Let the transition frequency (quantum beats frequency) be in resonance with the vibration frequency of the vibron ${\bf q}(t)$. Then for the direct transition $|1\rangle\to|2\rangle$ the vibron oscillation reduces the energy difference and increases the transition amplitude. When the system reaches the level $|2\rangle$, the vibron oscillates in the opposite direction and the energy level difference increases thereby decreasing the amplitude of the reverse transition. This effect can be compared with the formula (\ref{Amp}) where the value $\Delta$ is taken different for the forward and reverse transitions. In addition to resonance, it is also important to match the phases of the vibrons and quantum beats.
	
2) Tuning of decoherence. When the decoherence time coincides with the transition time, the forward transition works approximately in a coherent mode (hence with increased efficiency), and the reverse transition performs incoherently (hence with reduced efficiency), that increases the directionality of the transition. Such a regime can be described as collapse of the wave function at the moment of the transition.
	
Tuning both these transition parameters, i.e. transition resonance with the vibron and decoherence rate, has been discussed for quantum photosynthesis~\cite{Novoderezhkin2015}, \cite{Novoderezhkin2016}, \cite{Novoderezhkin2017}, \cite{Novoderezhkin2017_1}, where however a different model was used in comparison to the approach of this paper, in particular quantum description of the vibrons was applied. For the approach of this work, important is to match the phases of the vibron and quantum beats, which is done by using feedback in the system (see also the discussion below). That distinguishes this model from the approach of the aforementioned works. In \cite{Novoderezhkin2019} it was discussed that presence of mutations in proteins in the vicinity of the reaction center can reduce the transfer rate in quantum photosynthesis by two orders of magnitude. It can be conjectured that such mutations can affect the phase matching of the vibrons and quantum beats.

\subsection{Quantum photosynthesis}
	
\begin{figure}[t]
\center
\includegraphics[width = 0.4\linewidth]{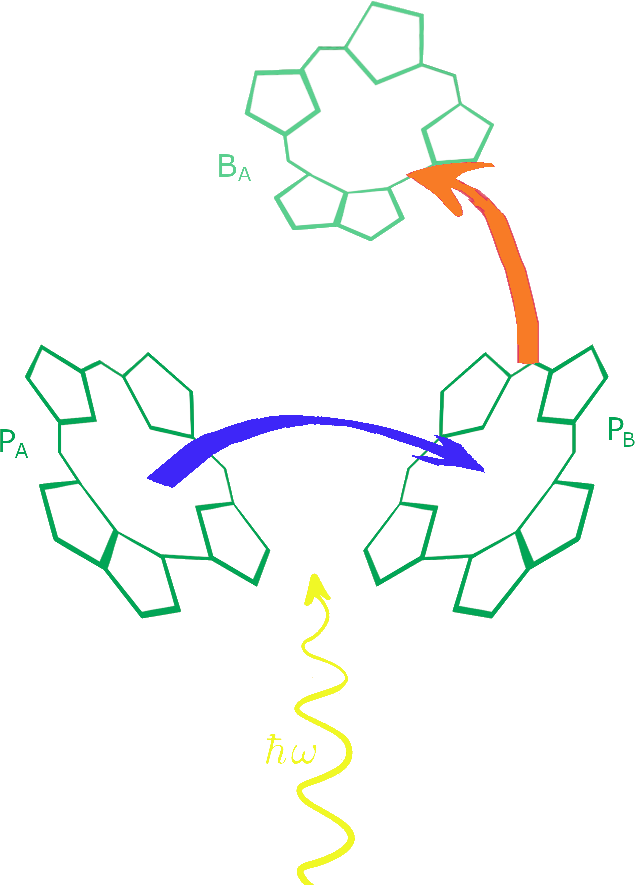}
\caption{Schematic picture of charge separation in the photosynthetic reaction center of the purple bacterium Rhodobacter (Rba.) sphaeroides. The reaction center contains a pair of strongly excitonically coupled bacteriochlorophyll $P_A$ and $P_B$, two monomeric BChl $B_A$ and $B_B$, two bacteriopheophytin $H_A$ and $H_B$ and two ubiquinone molecules symmetrically arranged in two branches ($A$ and $B$ branches; $A$ branch is not shown and only part of $B$ branch is shown; $B_B$, $H_A$, $H_B$ and two ubiquinone molecules are not shown). Charge separation (schematically shown by blue arrow) occurs in $P_A$ and $P_B$. Then the electron is transported to the monomeric BChl $B_A$ through the $A$ branch of the reaction center (shown by orange arrow).}
\label{FigRC}
\end{figure}

Schematic picture of charge separation in the photosynthetic reaction center is shown on figure~\ref{FigRC}. Chlorophyll contains magnesium atom, which is source of electron, and alternating single and double bonds (pi bond, or conductor). This  works as antenna in light harvesting complexes, where electronic excitations are created by absorption of photons. Also chlorophyll is a large molecule loosely bound (for photosynthetic complexes) to protein matrix. This allows existence of vibronic modes with comparably low energy (large wavelength) --- chlorophyll looks like a resonator for vibrons which can be in resonance with transition in the reaction center.
	
We make the following conjecture: high effectiveness of quantum photosynthesis is related to quantum ratchet which performs charge separation. To show this, we use the following parameters of the model taken from \cite{Novoderezhkin2016}. The Hamiltonian~(\ref{feedback}) takes the following form, where the sinus term describes the vibrons
\begin{equation}\label{model4}
H(t)=\left(\begin{array}{cc} 300 & 75 \cr 75 & 300\sin(340 t)\end{array}\right).
\end{equation}
Such a Hamiltonian is intended to describe photosystem II reaction center (PSII-RC), energy and vibron frequency are given in reciprocal centimeters (we measure frequencies, vibron amplitude $A= |{\bf v}\cdot {\bf w}|$, and energy in wavenumbers, that is, $\hbar=1$, $[\omega]=[A]=[E]=\textrm{cm}^{-1}$; here reciprocal centimeter is the energy unit used in spectroscopy which is the energy of light quantum with the wavelength one centimeter; it is approximately equal to  $1.24\times 10^{-4}~\textrm{eV}$). We take experimentally obtained values of energy difference $E_1-E_2=300~\textrm{cm}^{-1}$, electronic coupling $J=75~\textrm{cm}^{-1}$, and the vibron frequency $340~\textrm{cm}^{-1}$, respectively, from Table~1 in~\cite{Novoderezhkin2016}. Here the vibron frequency 340~$\textrm{cm}^{-1}$, was chosen because it corresponds to a known vibration of chlorophyll and was observed in the experimental 2D electronic spectroscopy spectra~\cite{RomeroNatPhys2014,FullerNatChem2014}.
	
The observed in experiment vibron frequency $340~\textrm{cm}^{-1}$ corresponds to almost exact resonance of quantum beats in Eq.~(\ref{Amp}),
$$
2\sqrt{(E_1-E_2)^2/4+J^2}=335\textrm{ cm}^{-1}.
$$
The decoherence time is chosen to be close to the transition half-period, so that the parameters in (\ref{theta}), (\ref{sink})  are
\begin{equation}\label{model41}
\gamma^{-}=60~\textrm{cm}^{-1},\quad \gamma^{+}/\gamma^{-}=0.22,
\end{equation}
where temperature $\beta^{-1}$ (which determines $\gamma^{+}/\gamma^{-}$ ratio) is room temperature (for room temperature $\beta^{-1}\approx 300~\textrm{K}\approx 200~\textrm{cm}^{-1}$ in reciprocal centimeters cm$^{-1}$), and the sink parameter $s_2=0.1$ is taken of order of magnitude slower compared to the transition $|1\rangle\rightarrow|2\rangle$ (both these choices are in agreement with the discussion of~\cite{Novoderezhkin2016}, where units non explicitly shown in the values of parameters $\gamma^{-}$ and $s_2$ are in hundreds of cm$^{-1}$).
	
For the quantum feedback model~(\ref{themaster1}) with the choice of the parameters as in~(\ref{model4}) we get for the Hamiltonian~(\ref{feedback1}) the expression
\begin{equation}\label{model5}
H(\rho(t),t)=\left(\begin{array}{cc} 300 & 75 \cr 75 & 300\left(\rho_{11}(t)+\rho_{22}(t)\right)\sin(340 t)\end{array}\right).
\end{equation}
	
\section{Simulations for optimization of the exciton recombination time}\label{Sec:Simulations}
	
\begin{figure}[t!]
\center
\includegraphics[width = 0.6\linewidth]{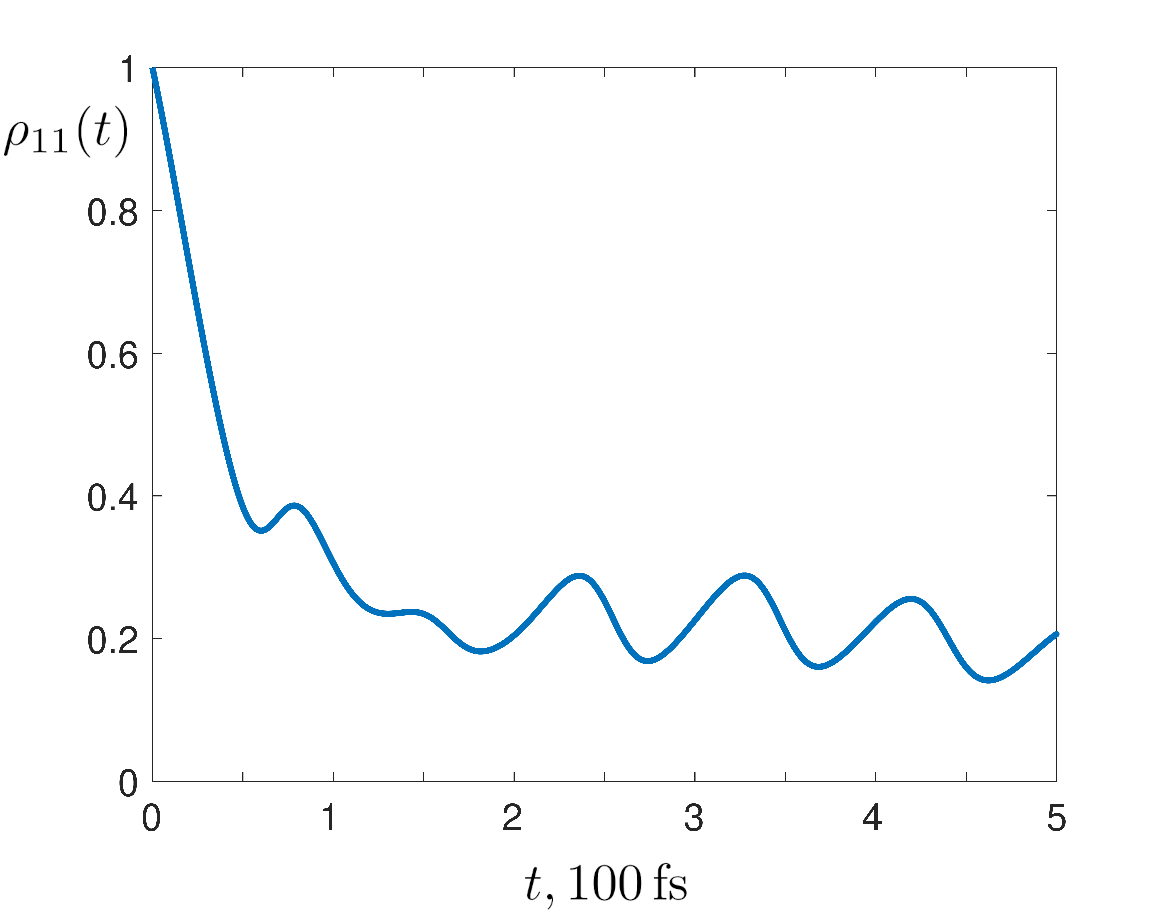}
\caption{Time evolution of $\rho_{11}$. The friction coefficient (dephasing rate) is $\gamma^{-}=60$ and the vibrons are $300\sin(340 t)$ (all values, e.g. amplitude, frequency and dephasing rate are in cm$^{-1}$).}
\label{Fig14}
\end{figure}

Simulation result for (\ref{themaster}) with the Hamiltonian~(\ref{model4}) for element $\rho_{11}(t)$ of the density matrix with the initial condition $\rho(0)=|1\rangle\langle 1|$ is shown in figure~\ref{Fig14}  (here $s_1=0$). The system performs a fast transition $|1\rangle\rightarrow|2\rangle$. Reverse transitions are suppressed. The directionality effect of the transition is important for the first two oscillations. Beyond the first two beats the system enters the regime of oscillations forced by the vibrons. Therefore integration in~(\ref{recombination}) should be restricted to a period approximately of two quantum beats.
	
\begin{figure}[t!]
\center
\includegraphics[width = \linewidth]{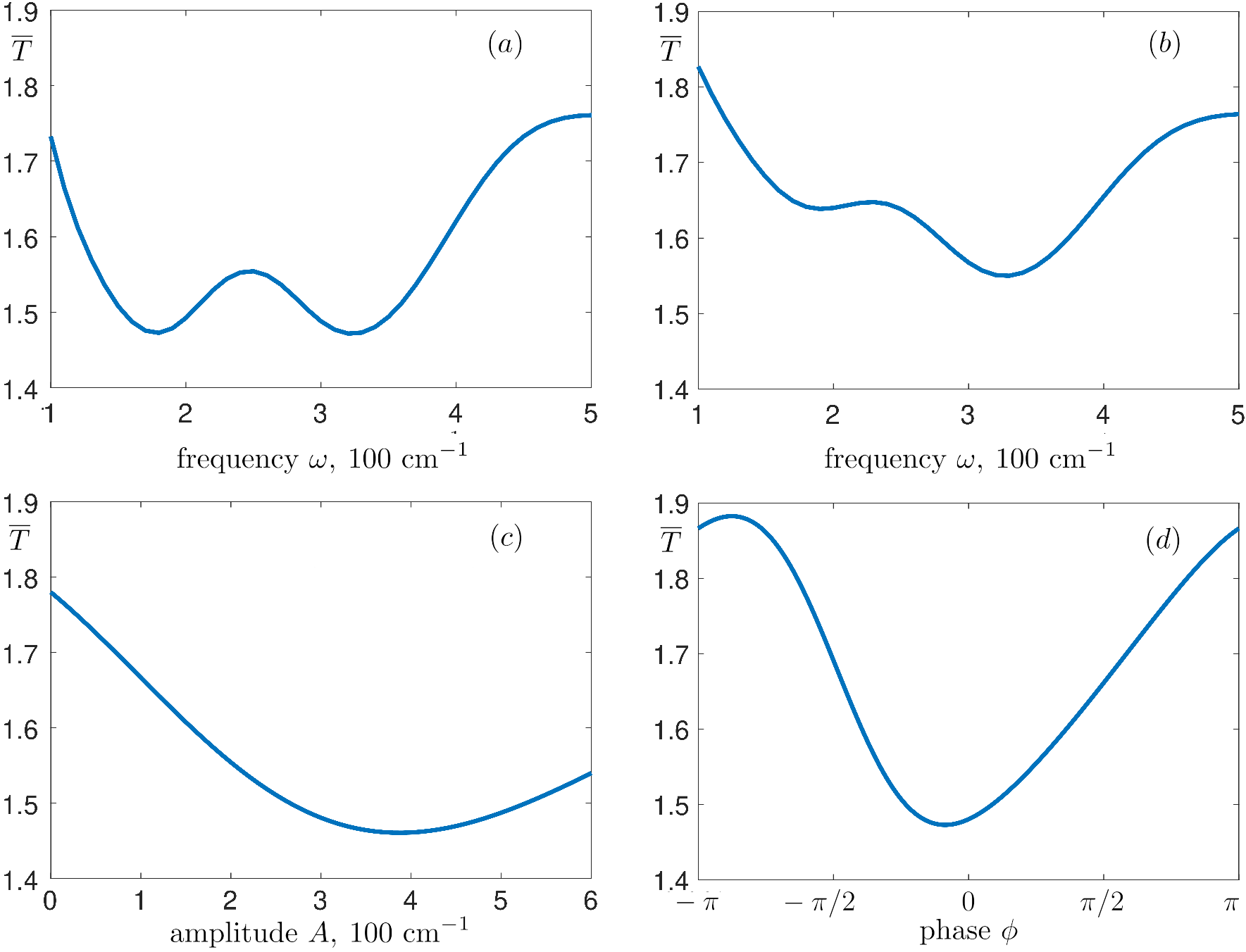}
\caption{Dependence of $\overline{T}=\int\limits_0^{T_0}\rho_{11}(t)\rmd t$ (in the units of $50~\textrm{fs}$) on various parameters of the vibron. Here $T_0=4\pi/3.40\times 50~\textrm{fs}\approx 185~\textrm{fs}$. Up: on the vibron frequency $\omega$ (phase $\phi=0$; left: amplitude $A=300~\textrm{cm}^{-1}$; right: amplitude $A=200~\textrm{cm}^{-1}$). Global minimum is small but present on the left and disappears on the right. Bottom left: on the vibron amplitude (frequency $\omega=340~\textrm{cm}^{-1}$). Bottom right: on the vibron phase (amplitude $A=300~\textrm{cm}^{-1}$, frequency $\omega=340~\textrm{cm}^{-1}$).}
\label{Fig4}
\end{figure}

To understand better operation of the quantum ratchet we consider numerical simulation of the dependence of $\overline{T}=\int\limits_0^{T_0} \rho_{11}(t)\rmd t$ for $T_0=2\times 2\pi/3.40\times 50\approx 185$~fs on various parameters of the system. Here the upper time limit of 185~fs in the integral corresponds to a period approximately of two quantum beats of the transition, see figure~\ref{Fig14}. We plot the dependence of $\overline{T}$ on various parameters of the vibron in the subfigures in figure~\ref{Fig4}.
	
In the upper row [figure~\ref{Fig4}(a) and figure~\ref{Fig4}(b)], dependence of $\overline{T}$ on frequency of the vibron is shown. On the left upper subfigure [figure~\ref{Fig4}(a)] the vibron amplitude is 300 cm$^{-1}$. Two minima of $\overline{T}$ are observed, the first at frequency about $170~\textrm{cm}^{-1}$ and the second at frequency about $330~\textrm{cm}^{-1}$ which is close to the experimentally observed value $340~\textrm{cm}^{-1}$. The first minimum is an artifact of integration over the restricted time period, since it vanishes with changing the time period or with decreasing the vibron amplitude. To compare, in the right subfigure [figure~\ref{Fig4}(b)] the same dependence is shown for the vibron amplitude $200~\textrm{cm}^{-1}$, for which the minimum at frequency $170~\textrm{cm}^{-1}$ almost disappears which suggests that this minimum might be not algorithmically stable, see the discussion below. Whether the minimum at frequency $170~\textrm{cm}^{-1}$ can or can not be presented in experimental situations is an open question.
	
In the bottom row, left subfigure [figure~\ref{Fig4}(c)], we show the dependence of  $\overline{T}$ on the amplitude of the vibron in the interval $[0,600~\textrm{cm}^{-1}]$ (frequency $\omega=340~\textrm{cm}^{-1}$, phase $\phi=0$, other parameters are as above). We observe the minimum of $\overline{T}$ between the amplitudes $300~\textrm{cm}^{-1}$ and $400~\textrm{cm}^{-1}$.
	
In the bottom row, right subfigure [figure~\ref{Fig4}(d)], we show the dependence of $\overline{T}$ on the phase $\phi$ of the vibron (amplitude $A=300~\textrm{cm}^{-1}$, frequency $\omega=340~\textrm{cm}^{-1}$). The minimum of $\overline{T}$ is observed at the phase value $\phi\approx-0.2$.
	
\begin{figure}[t!]
\center
\includegraphics[width = 0.6\linewidth]{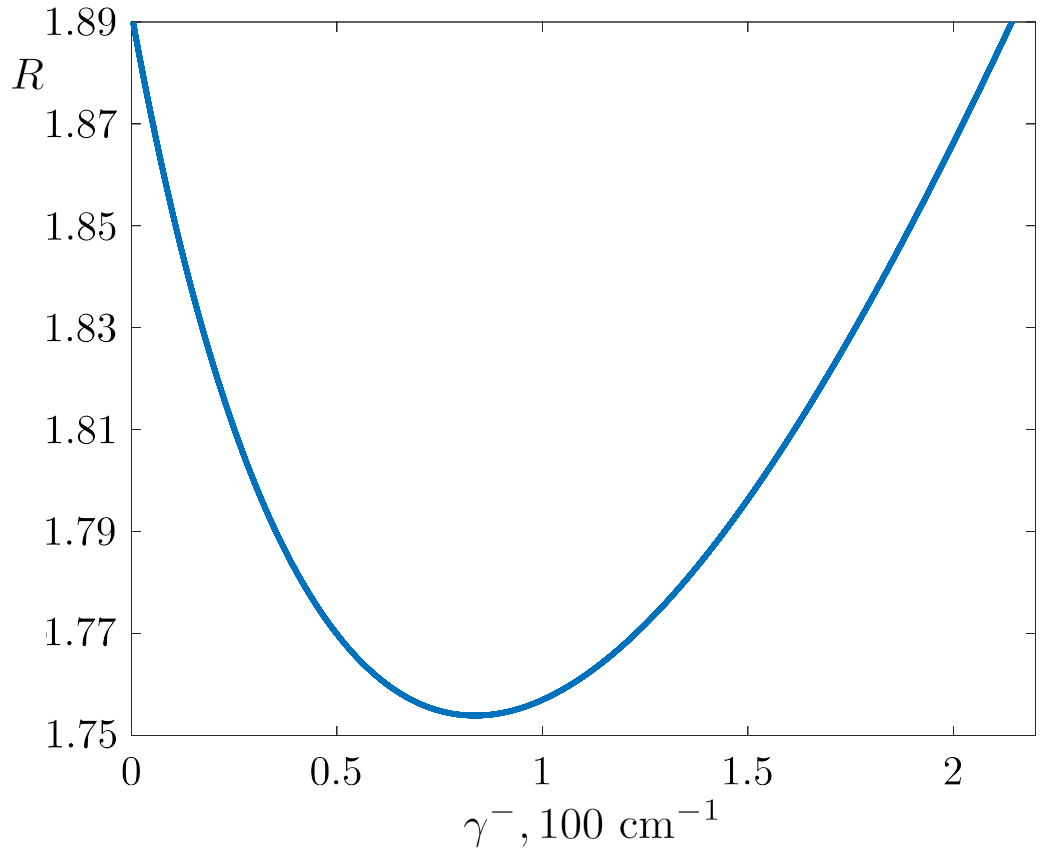}
\caption{Dependence of $R=(50~\textrm{cm}^{-1}+0.22\gamma^{-})\overline{T}/\hbar$ on the friction $\gamma^{-}$. The amplitude is $300~\textrm{cm}^{-1}$, the frequency is $340~\textrm{cm}^{-1}$, and the phase is zero.}
\label{Fig35}
\end{figure}

It is natural to assume that the exciton recombination velocity increases with the dephasing rate $\gamma^{-}$. We propose to take the recombination proportional to the dimensionless quantity $R=\Bigl(50~\textrm{cm}^{-1}+0.22\gamma^{-}\Bigr)\int\limits_0^{185\,[\textrm{fs}]} \rho_{11}(t)\rmd t/\hbar$. In figure~\ref{Fig35} we show the dependence of this value on the friction parameter $\gamma^{-}$ for amplitude $A=300~\textrm{cm}^{-1}$, frequency $\omega=340$ cm$^{-1}$, and phase $\phi=0$. Its minimum is achieved at $\gamma^{-}\approx 70~\textrm{cm}^{-1}$.
	
In the discussed above sense the parameters of the model given in (\ref{model4}), (\ref{model41}) can be considered close to optimal --- these parameters give with a good accuracy the minimum of recombination of excitons at the time interval $[0,185~\textrm{fs}]$. The obtained optimal parameters are close to experimentally measured values provided in Table~1 in~\cite{Novoderezhkin2016}.
	
\begin{figure}[t]
\center
\includegraphics[width = \linewidth]{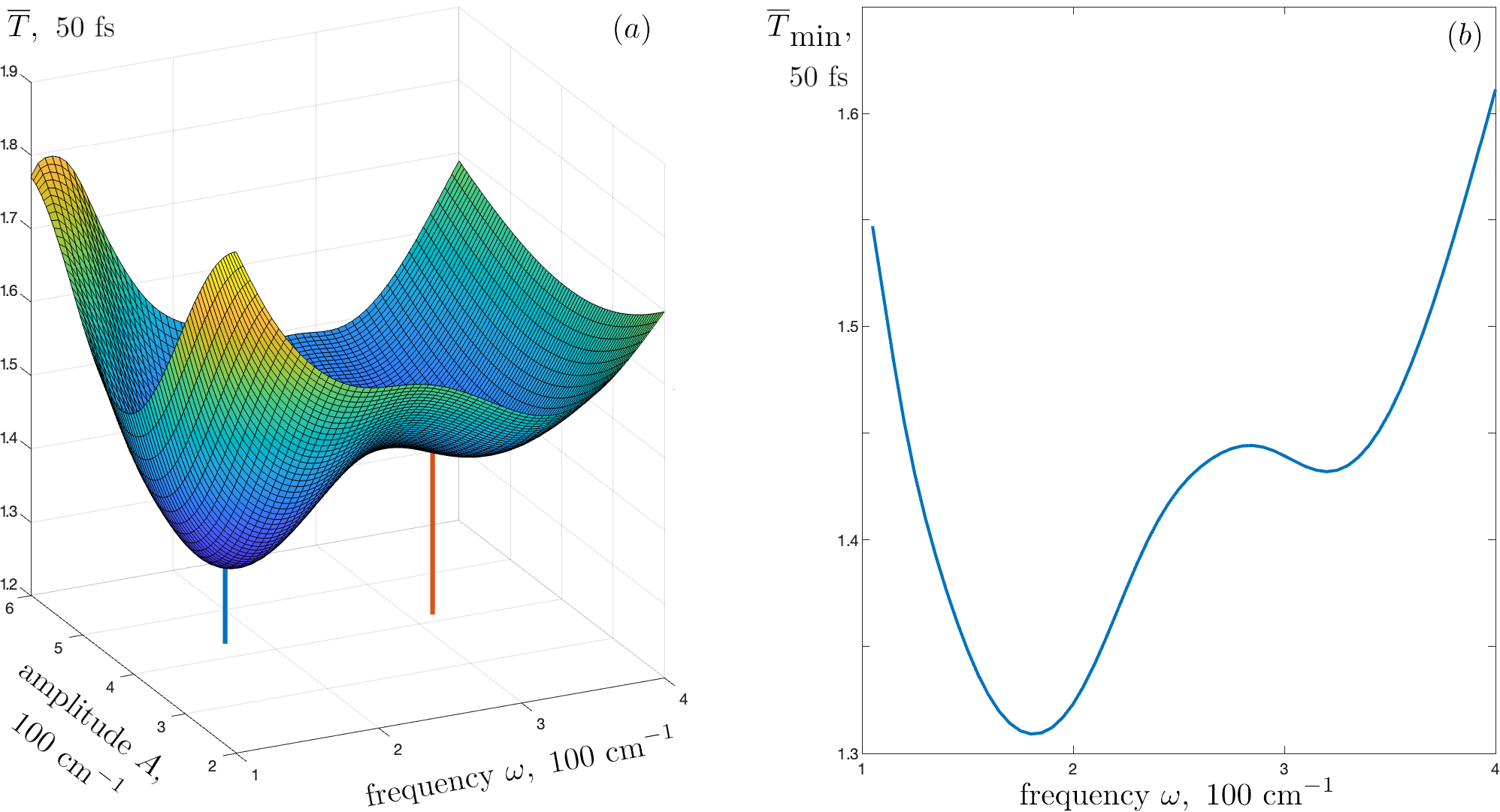}
\caption{Left: Average population $\overline T(A,\omega)=\int\limits_0^{T_0}\rho_{11}(t)\rmd t$ as a function of the amplitude $A$ and frequency $\omega$ of the vibron. Right: minimal value of $\overline T(A,\omega)$ over $A$ for each $\omega$, i.e., $\overline{T}_{\min}(\omega):=\min\limits_A \overline T(A,\omega)$. Here $\gamma^-=50~\textrm{cm}^{-1}$, $\gamma^+=0.22\gamma^-$, $E_1-E_2=300~\textrm{cm}^{-1}$, $J=75~\textrm{cm}^{-1}$, and $T_0=4\pi/ 3.35\times 50~\textrm{fs}\approx 185~\textrm{fs}$. On the left 3D plot, the red line shows position of the local minimum, which corresponds to experimentally known parameters, and the blue line shows position of the global minimum.}
\label{Fig3_3D}
\end{figure}
	
On figure~\ref{Fig3_3D} we show the behaviour of $\overline T=\overline T(A,\omega)$ plotted as a function of both the amplitude $A$ and frequency $\omega$ of the vibron. We take the parameters $\gamma^-=50~\textrm{cm}^{-1}$, $\gamma^+=0.22\gamma^-$, $E_1-E_2=300~\textrm{cm}^{-1}$, $J=75~\textrm{cm}^{-1}$, and $T_0=2\times 2\pi/3.35\times 50\approx 185~\textrm{fs}$ which is approximately a period of two quantum beats.
	
We discover two minima, local and global, of the average population. The local minimum is obtained at $A=415~\textrm{cm}^{-1}$ and $\omega=315~\textrm{cm}^{-1}$ with the value
$\overline T\approx 71~\textrm{fs}$. Its position is shown by the red line on the figure~\ref{Fig3_3D}(a). The corresponding frequency is close to the frequency of quantum beats, although a bit smaller. So the period of the vibron is a bit more than the period of one quantum beat. The global minimum is obtained at $A=430$ cm$^{-1}$ and $\omega=175$ cm$^{-1}$ with the value $\overline T\approx 65~\textrm{fs}$. Its position is shown by the blue line on figure~\ref{Fig3_3D}(a). In this case frequency is about half of frequency of the quantum beat, and period in opposite, it about two periods of quantum beat. Important is that global minimum is separated from the local minimum by a barrier and can not be achieved by small local changes of the parameters of the vibron. The figure~\ref{Fig3_3D}(b) shows minimal value of $\overline T(A,\omega)$ over $A$ for each $\omega$, i.e. shows the dependence of $\overline T_{\rm min}(\omega):=\min\limits_A \overline T(A,\omega)$ on the vibron frequency $\omega$. Local and global minima are clearly visible. Therefore our model predicts, in addition to existing, a frequency about half of frequency of the quantum beat, and period in opposite, of about two periods of quantum beat. As discussed above, the left minimum at figure~\ref{Fig3_3D}(a) (shown by the blue line) disappears with decreasing of the vibron amplitude and is a corollary of the chosen optimization procedure which involves large amplitudes of the vibron. This prediction of the model might be investigated in experiments.
	
Summing up, we conjecture that the model of operation of the photosynthetic reaction as a quantum ratchet explains high performance of quantum photosynthesis. Simulation with quantum feedback with the Hamiltonian (\ref{model5}) instead of (\ref{model4}) does not show a considerable improvement in performance of quantum transport. For the effect of quantum ratchet, tuning phases of the vibrons and quantum beats is important. While this tuning is done in (\ref{model4}) by hands, it is related to the quantum feedback mechanism in~(\ref{model5}).
	
\section{Discussion}\label{Sec:Discussion}
	
\begin{remark}
In a (semiclassical) model of laser \cite{Haken} resonance of the laser mode with the energy difference of transition between levels coupled to the laser mode works as an amplifier of the quantum transfer at this transition. The described above quantum ratchet works in a similar way, but only for the direct transition $|1\rangle\rightarrow|2\rangle$ (where coupling to the vibron reduces the energy difference for the transition). For the reverse transition $|2\rangle\rightarrow|1\rangle$ the vibron oscillation increases this energy difference and slows down the transition which gives the ratchet effect. This effect is achieved by tuning phase of quantum beats and vibrons coupled to the transition.
\end{remark}
	
\begin{remark}
In \cite{Novoderezhkin2021} Stark effect in quantum photosynthesis is discussed. It is shown that Stark effect shifts energies of charge separation states (states $|1\rangle$
and $|2\rangle$ in the notations of our paper) and makes the transition to the next state in the charge transfer chain more effective. This transition is described by Marcus theory. The Stark effect allows to perform tuning of the transition parameters to make these parameters optimal. This can be considered as a control which indicates that control-like effects appear in several steps of quantum photosynthesis.
\end{remark}
	
\begin{remark}
Simulation shows that slight detuning of the parameters of photosynthetic center (such as frequency and amplitude of the vibrons, as well as decoherence rate) from the optimal parameters given in figure \ref{Fig14} does not change drastically the transfer efficiency (i.e. the minimum of $\overline{T}$ in the space of these parameters is flat). Therefore the problem of finding optimal parameters of photosynthetic center is algorithmically stable from the point of view of learning theory \cite{Poggio}. This improves the solvability of the problem of evolution of photosynthetic reaction centers.
\end{remark}
	
Summing up, in the present paper we discuss vibrons at the photosynthetic reaction center as a model of quantum control which describes the quantum ratchet --- a model of quantum technology performing directed transitions in quantum transport. Amplitude and frequency of the oscillating vibron together with the dephasing rate are the parameters of the quantum ratchet which determine its efficiency.  We adjust the parameters of the quantum ratchet to minimize the exciton recombination time which is proportional to the effective time during which the quantum system stays in the excited state for a period of two quantum beats. We find optimal values of the parameters of the quantum ratchet --- such that with these values of the parameters the quantum ratchet minimizes the exciton recombination time. We make simulations which show that experimentally observed values of the parameters of the photosynthetic reaction center correspond to these optimal values of the parameters of the quantum ratchet. We also show that these optimal values of the parameters realize a minimum which is sufficiently wide and flat. That implies the algorithmic stability for evolution of photosynthetic reaction center. We also find another parameters of the quantum ratchet realizing another minimum of the excition recombination time which corresponds to a twice smaller frequency of the vibron compared to that observed in experiments. Whether it can or can not be presented in experimental situations of photosynthesis is an open question. Beyond photosynthesis and without bounds on the amplitudes, it can play the role for optimization of quantum transport.
\bigskip

\noindent {\bf Acknowledgments}. This work was funded by the Ministry of Science and Higher Education of the Russian Federation (grant number 075-15-2020-788).

\end{document}